\documentclass[12pt,preprint]{aastex}
\usepackage{graphicx}
\usepackage{natbib}

\def\ensuremath#1{{\ifmmode #1 \else $#1$\fi}}
\newcommand{\msun}{\ensuremath{\mathrm{M_{\sun}}}}

\begin{document}
\title{A New $^{17}$F(p,$\gamma$)$^{18}$Ne Reaction 
Rate and Its Implications for Nova Nucleosynthesis}

\author{
S. Parete-Koon\altaffilmark{1,2},
W. R. Hix\altaffilmark{1,2,3},
M.S. Smith\altaffilmark{2},
S. Starrfield\altaffilmark{4},
D.W. Bardayan\altaffilmark{2},
M.W. Guidry\altaffilmark{1,2},
A. Mezzacappa\altaffilmark{2}
}

\altaffiltext{1}{Department of Physics \& 
Astronomy, University of Tennessee,
Knoxville, Tennessee 37996-1200}

\altaffiltext{2}{Physics Division, Oak Ridge 
National Laboratory, Oak Ridge, 
TN 37831-6354}

\altaffiltext{3}{Joint Institute For Heavy Ion 
Research, Oak Ridge National Laboratory,
Oak Ridge, Tennessee 37831-6374}

\altaffiltext{4}{Department of Physics \& 
Astronomy, Arizona State University,
Tempe, AZ 85287-1504}

\begin{abstract}

%\tighten

Proton capture by $^{17}$F plays an important role in the synthesis of
nuclei in nova explosions. A revised rate for this reaction, based on a
measurement of the $^1$H($^{17}$F,p)$^{17}$F excitation function using a
radioactive $^{17}$F beam at ORNL's Holifield Radioactive Ion Beam
Facility, is used to calculate the nucleosynthesis in nova outbursts on the
surfaces of 1.25 $\msun$ and 1.35 $\msun$ ONeMg white dwarfs and a 1.00
$\msun$ CO white dwarf. We find that the new $^{17}$F (p,$\gamma$)
$^{18}$Ne reaction rate changes the abundances of some nuclides (e.g.,
$^{17}$O) synthesized in the hottest zones of an explosion on a 1.35
$\msun$ white dwarf by more than a factor of 10$^4$ compared to
calculations using some previous estimates for this reaction rate, and by
more than a factor of 3 when the entire exploding envelope is considered.
In a 1.25 $\msun$ white dwarf nova explosion, this new rate changes the
abundances of some nuclides synthesized in the hottest zones by more than a
factor of 600, and by more than a factor of 2 when the entire exploding
envelope is considered. Calculations for the 1.00 $\msun$ white dwarf nova
show that this new rate changes the abundance of $^{18}$Ne by 21\%, but has
negligible effect on all other nuclides. Comparison of model predictions
with observations is also discussed.
\end{abstract}


\section{Introduction}

Nova explosions result from the transfer of stellar material onto a white
dwarf star (WD) from a  companion star. The mass transfer and resulting
rise in temperature initiate hydrogen burning via the CNO cycle and trigger
a violent thermonuclear explosion on the WD surface providing an energy
release of up to $\sim$ 10$^{46}$ ergs with the peak of the thermonuclear
runaway lasting up to 1000 seconds and the observed outburst lasting years
\citep{GTWS98}. These outbursts are the largest hydrogen driven
thermonuclear explosions in the Universe, and are characterized by high
temperatures and densities in the nuclear burning region - greater than
10$^8$ K and 10$^4$ g cm$^{-3}$, respectively \citep{StTS78}.

Under such conditions, proton and $\alpha$ particle capture reactions on
proton-rich radioactive nuclei become faster than  $\beta$$^+$ decays.
Unstable nuclei produced by capture reactions can then undergo further
reactions before they decay, resulting in a sequence of reactions (the
rapid proton capture process, or $rp$-process) that is very different from
the sequences in non-explosive environments \citep{WaWo81}.  This explosive
hydrogen burning generates energy up to 100 times faster than in the
quiescent burning phase and drives the outburst.  The timescale for these
nuclear reactions is comparable to that of convection in the exploding
envelope \citep{Star89,ShAr94}, allowing the mixing of unstable nuclei into
the outer envelope of the nova.  For these reasons, accurate determinations
of the rates of reactions on proton-rich radioactive nuclei are vitally
important to our understanding of these explosions \citep{WiSc98,JoCH99}.

In addition to driving the outburst, nuclear reactions in novae can
synthesize nuclides up to mass $A \sim 40$, producing an abundance pattern
distinct from that in CNO burning in quiescent stars \citep{VSWS96,CHLM99}.
 For example, long-lived radioactive nuclei such as $^{18}$F are
synthesized and ejected. Because of its relatively long half-life and
significant abundance, it has been suggested that the decay of $^{18}$F in
the nova ejecta produces the majority of observable gamma rays during the
first several hours after the explosion \citep{HTWP01}. The observation of
such gamma rays may provide a rather direct test of nova models
\citep{LeCl87,HNTC99, HJCG99}. However, the quantity of $^{18}$F produced in
the interior and transported to the top of the nova envelope is severely
constrained by the nuclear reactions that create and destroy $^{18}$F. The
sensitivity required to make gamma-ray observations with orbital detectors
\citep[e.g., INTEGRAL, see][]{HGJC01} is therefore impossible to determine
without a better understanding of the reactions that create and destroy
$^{18}$F. Whereas the $^{18}$F(p,$\alpha$)$^{15}$O reaction rate is the
primary destruction mechanism of $^{18}$F \citep{CHJT00}, $^{18}$F can be
produced via two different reaction sequences:
$^{17}$F(p,$\gamma$)$^{18}$Ne$(\beta$$^+$$\nu$)$^{1 8}$F and
$^{17}$O(p,$\gamma$)$^{18}$F. We pay particular attention to the changes in
$^{18}$F production arising from a newly calculated
$^{17}$F(p,$\gamma$)$^{18}$Ne reaction rate.

\section {The $^{17}$F(p,$\gamma$)$^{18}$Ne Reaction}

\cite{WiSc98} suggested that the $^{17}$F(p,$\gamma$)$^{18}$Ne reaction
rate could influence the amount of  $^{15}$O, $^{17}$F,  $^{18}$Ne, and
$^{18}$F produced in novae. This reaction is also part of a sequence of
reactions providing a possible transition from hot CNO cycle burning to
the $rp$-process in the most energetic novae \citep{WaWo81}. An unnatural
parity (J$^{\pi}$ = $3^+$) state in $^{18}$Ne provides an $\ell = 0$
resonance in $^{17}$F + p capture which was thought to dominate the
$^{17}$F(p,$\gamma$)$^{18}$Ne reaction rate at nova temperatures
\citep{WiGT88}. This level was expected from the structure of the $^{18}$O
isobaric mirror nucleus, but never conclusively observed despite nine
experimental studies of the relevant excitation energy region in $^{18}$Ne.
Different determinations of the properties (excitation energy, total width)
of this level - based on shell model calculations - resulted in differences
of more than a factor of 100 in the $^{17}$F(p,$\gamma$)$^{18}$Ne reaction
rate \citep{WiGT88,GAMM91,ShFo98}. The rate used in REACLIB
\citep{Reaclib95}, the reaction rate library most widely-used for nova
simulations, is that of \cite{WiGT88}, which is the fastest rate primarily
because their estimate of the excitation energy for the J$^{\pi}$ = $3^+$
$^{18}$Ne resonance was the lowest.

\begin{figure}
    \includegraphics[width=\linewidth] {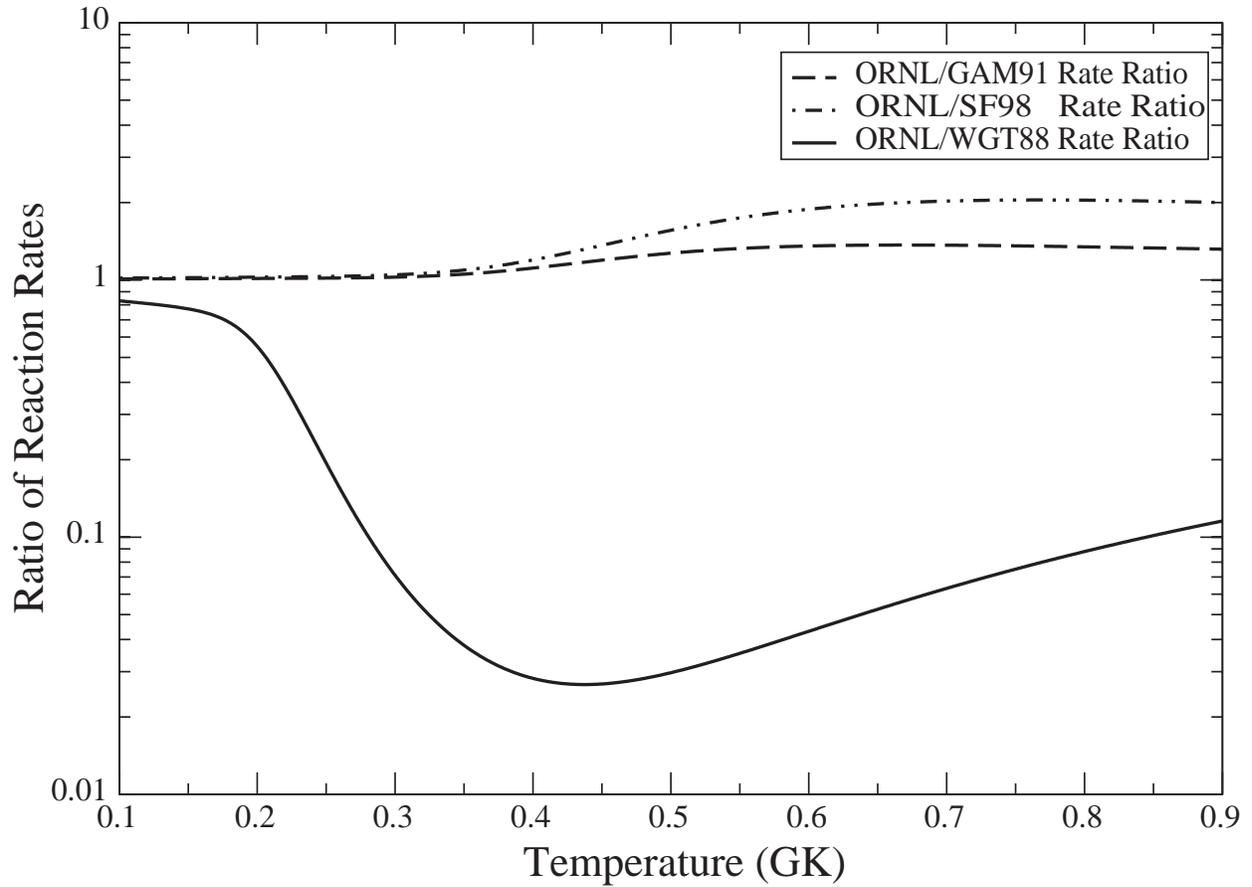}
    \caption{Ratio of the new ORNL $^{17}$F(p,$\gamma$)$^{18}$Ne rate   
    \citep{BBBC00} to three previous estimates, WGT88  \citep{WiGT88}, 
    SF98 \citep{ShFo98}, and GAM91 \citep{GAMM91}, as a function of 
    stellar temperature. The temperature range of interest for novae is 
    approximately 0.1-0.4 GK. \label{fig:rateratio}} 
\end{figure}

A measurement of the excitation function for the $^1$H($^{17}$F,p)$^{17}$F
reaction at ORNL's Holifield Radioactive Ion Beam Facility (HRIBF) was used
to obtain the first unambiguous evidence for the J$^{\pi}$ = $3^+$ state in
$^{18}$Ne and precisely determine its energy and total width
\citep{BBBC99}. Using a calculation of the (still unmeasured)
$\gamma$-width of this level, its contribution to the
$^{17}$F(p,$\gamma$)$^{18}$Ne reaction rate \citep{BBBC00} was determined,
thereby resolving the greatest uncertainty in the $^{17}$F(p,$\gamma$)$^{18}$Ne reaction rate. Combining this measurement
with calculations of the non-resonant direct capture rate from 
\cite{GAMM91}, the $J^{\pi}$ = $3^+$ level is now estimated to only dominate the capture rate at temperatures above 5 $\times$ 10$^8$ K, while direct capture dominates at the lower temperatures ($\lesssim 4 \times 10^8$ K) characteristic of novae. A parameterization of this reaction rate is given in the REACLIB format in \cite{BBBC00}. To explore the astrophysical 
impact of the ORNL measurement \citep{BBBC99}, which only influences the 
resonant reaction rate, we constructed four different reaction rates for our study - based on the direct capture rate from GAM91 added to the resonant 
rate using parameters from \cite{WiGT88,GAMM91,ShFo98,BBBC99}. Hereinafter, we will refer to these as the WGT88, GAM91, SF98, and ORNL rates, respectively. The ratios of the ORNL $^{17}$F(p,$\gamma$)$^{18}$Ne reaction 
rate to the other three rates are shown in Figure~\ref{fig:rateratio}.
At nova temperatures ($\sim 0.1-0.4$ GK), the new ORNL rate
differs only slightly from the SF98 and GAM91 rates, but differs by up to a
factor of 30 from the rate based on the WGT88 resonance 
parameters. These latter parameters are the basis for the 
rate in the widely-used REACLIB reaction rate library. 

\section{Nova Nucleosynthesis Calculations}

The temporal evolution of the isotopic composition in these nova explosions
was followed using a nuclear reaction network \citep{HiTh99b} containing
169 isotopes, from hydrogen to $^{54}$Cr with nuclear reaction rates drawn
from REACLIB \citep{Reaclib95}. In this paper we examine the
nucleosynthesis of 3 models for nova explosions, on a 1.00 $\msun$ CO WD,
on a 1.25 $\msun$ ONeMg WD, and on a 1.35 $\msun$ ONeMg WD. The first two
are representative of the most prevalent classes of novae, while the third
represents a more energetic outburst. Thirty to fifty per cent of all novae
are thought to occur on ONeMg white dwarfs \citep{GTWS98,GGJH03}.

In many prior nucleosynthesis calculations, the nuclear reaction network
was evolved under conditions of constant temperature and density
\citep[e.g.,][]{ChWi92,VGIW94}. This neglects the strong coupling of the
nucleosynthesis to the hydrodynamics. This coupling is important because
the nuclear reactions generate the energy powering the outburst, and the
reaction rate between any two nuclear species is highly variable in time
due to its dependence on the temperature and density. A fully consistent
description of an outburst therefore involves the coupling of a large
reaction network with a multidimensional hydrodynamic calculation of the
outburst, an approach which is made computationally demanding by the
similar timescales of the nuclear reactions and convective motions in
novae.  However, such simulations are in their infancy
\citep{ShAr94,KeHT99}.

We have improved on constant temperature and density nucleosynthesis
calculations by extracting hydrodynamic trajectories -- time histories of
the temperature and density -- from one-dimensional hydrodynamic
calculations for outbursts on 1.0, 1.25, and 1.35 $\msun$ white dwarfs
\citep[similar to those of][]{STWS98} which employed a more limited
reaction rate network. Different mass elements (\emph{zones}) of the
envelope at different radii generate unique trajectories. For example, the
temperature history of the hottest zone of the 1.25 $\msun$ ONeMg WD nova
appears later in Figure~\ref{fig:yvst}. In our simulations, the ejecta of
each of the nova models consisted of between 26 and 31 zones. Separate
\emph{post-processing} nuclear reaction network calculations with the full
complement of nuclei and nuclear reactions were carried out to study the
nucleosynthesis details within each zone; no mixing between the zones was
included. To calculate the total abundances in the ejecta of each
explosion, a sum was made of abundances over the zones, weighted by the
ratio of the zone mass to the total envelope mass. It should be noted that 
calculations of nova outbursts on similar WD projenitors 
carried out by different groups \citep[e.g.,][]{JoCH99,WaHN99} have yielded 
moderately different hydrodynamic trajectories (and peak temperatures) .  

The calculations for the 1.25 $\msun$ and 1.35 $\msun$ ONeMg WD novae begin
with a set of initial abundances adopted from \cite{PSTW95}.  They assumed
a solar composition mixed equally (by mass) with the ashes of carbon
burning (50\% $^{20}$Ne, 30\% $^{16}$O, and 20\% $^{24}$Mg). The initial
composition for the 1.00 $\msun$ CO WD nova was 50\% solar \citep{AnGr89}
(by mass) and 50\% products of He burning (an equal mix of $^{12}$C and
$^{16}$O with a trace of $^{22}$Ne). The composition in each case is
representative of the envelope material mixing with the matter from the
underlying white dwarf \citep{StST74}.

To see the effect of the $^{17}$F(p,$\gamma$)$^{18}$Ne reaction rate on the
nucleosynthesis, this post-processing nucleosynthesis calculation was run
for each nova model with each of the reaction rates, those based on 
resonance parameters from ORNL \citep{BBBC99}, WGT88 \citep{WiGT88}, 
SF98 \citep{ShFo98}, and GAM91\citep{GAMM91}, substituted into the reaction 
rate library. In each calculation, the only reaction rates changed in the library are the $^{17}$F(p,$\gamma$)$^{18}$Ne reaction rate and its inverse (obtained via detailed balance). At late times in the outburst, the
adiabatic expansion drops the temperatures of the ejecta below $10^7$ K,
where only the weak reactions significantly change the nuclear abundances.
Additionally, the reaction rate parameters compiled in REACLIB
\citep{Reaclib95} are valid only from 10$^7$ K to $10^{10}$ K. For these
reasons, while the complete set of reactions was used to evolve the
abundances for temperatures in excess of $10^7$ K, only the weak reactions
were used at lower temperatures. The simulations were stopped one hour
after the peak temperature was reached in the hottest zone so that the
potentially observable abundances of long-lived radionuclides (e.g.,
$^{18}$F) could be determined. Our analysis of the impact of the new
$^{17}$F(p,$\gamma$)$^{18}$Ne reaction rate was, however, insensitive to
the stop time.

For each nova, the abundance pattern produced by the network calculations
with the ORNL $^{17}$F(p,$\gamma$)$^{18}$Ne rate was compared to the
abundance pattern produced by the network calculations done with the other
$^{17}$F(p,$\gamma$)$^{18}$Ne rates. This analysis was performed for each
zone individually and for a weighted sum of all zones. Particular scrutiny
was given to the influence of the new $^{17}$F(p,$\gamma$)$^{18}$Ne
reaction rate on the production of the potentially observable radioisotope
$^{18}$F. Careful attention was also paid to the changes in nuclear energy
generation resulting from changing the $^{17}$F(p,$\gamma$)$^{18}$Ne
reaction rate. The post-processing approach used to calculate the
nucleosynthesis is valid as long as changes to the reaction rate library
result in negligible changes in the energy production, and therefore would
not alter the temperature and density history of the explosion. The
hydrodynamic profiles were generated by a network which used the WGT88 rate
for the $^{17}$F(p,$\gamma$)$^{18}$Ne rate. We compared the energy
generation of the calculations with the ORNL, SF98 and GAM91 rates to the
energy generation by the calculation with the WGT88 rate. For the 1.25
$\msun$ WD nova, the calculations with the ORNL, SF98 and GAM91 rates
showed less than 1\% difference in energy generation from the calculation
with the WGT88 rate for the entire exploding envelope and 2.13\%, 2.21\%,
and 2.18\% difference, respectively, for the hottest zone. For the 1.35
$\msun$ WD, the ORNL, SF98, and GAM91 rates resulted in energy generation
differences of 1.02\%, 1.10\%, and 1.06\% for the sum of all zones and
0.02\%, 0.08\%, and 0.05\%, respectively, for the hottest zone compared to
energy generation resulting from the WGT88 rate. There was less than 0.01\%
difference between the energy generation for rates in the 1.00 $\msun$ WD
nova calculations. From these results, we conclude that our rate variations
cause a negligible change in the temperature and density of the explosion,
verifying the validity of our post-processing nucleosynthesis calculations.

\section{Results for a 1.25 $\msun$ ONeMg WD Nova}

\begin{figure}
    \includegraphics[width=\linewidth] {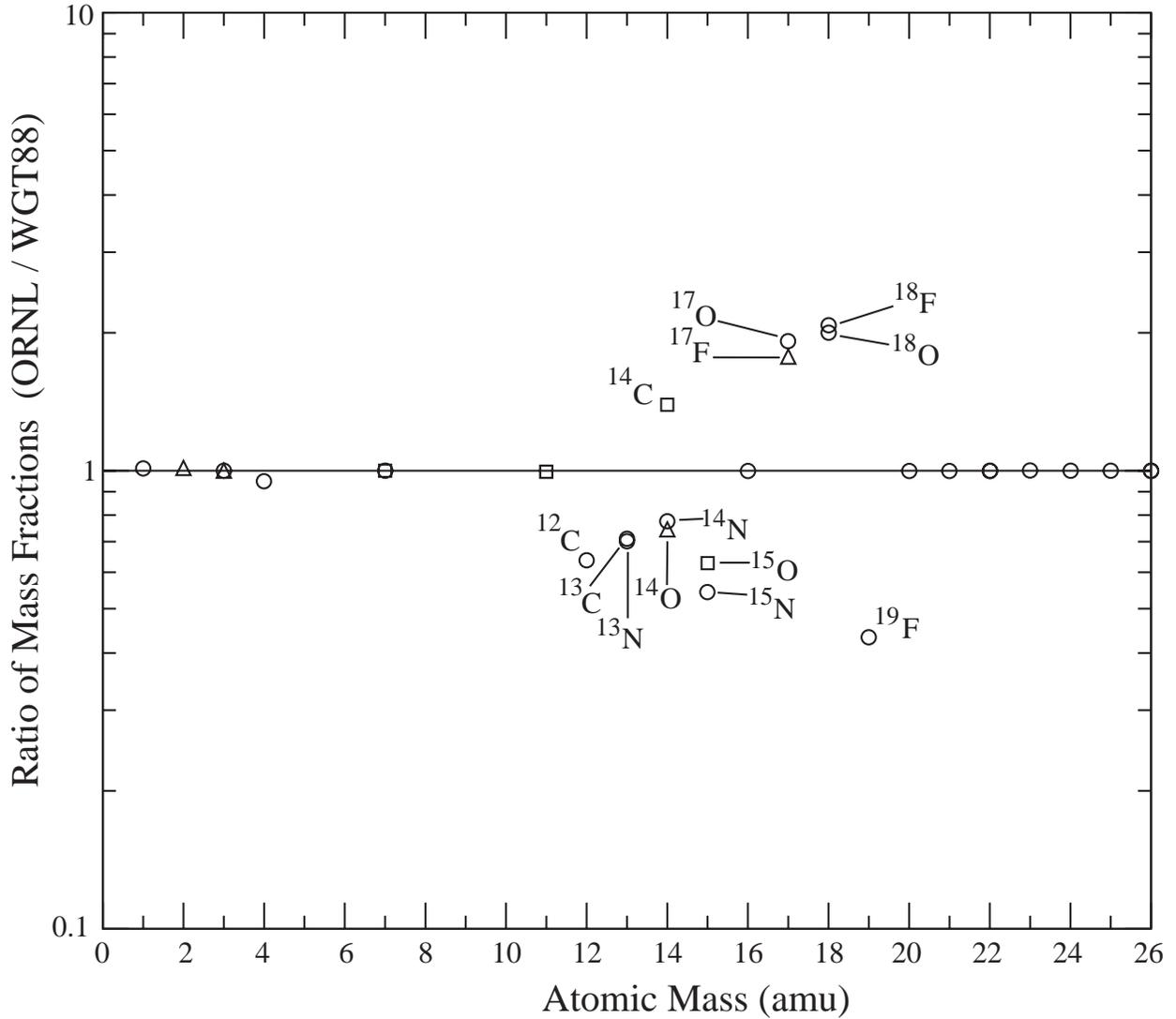} 
    \caption{The ratio of mass fractions (ORNL/WGT88) plotted against   
    nuclide mass for the entire envelope of a 1.25 $\msun$ WD nova. The 
    ORNL rate changes the mass fractions of some nuclei by up to a  
    factor of 2. The circular symbols mark species with mass fractions 
    greater than 10$^{-8}$, the square symbols mass fractions between 
    10$^{-8}$ and 10$^{-16}$, and the triangular symbols mass fractions 
    less than 10$^{-24}$. 
    \label{fig:ratio125}}
\end{figure}

\begin{figure}
    \includegraphics[width=\linewidth] {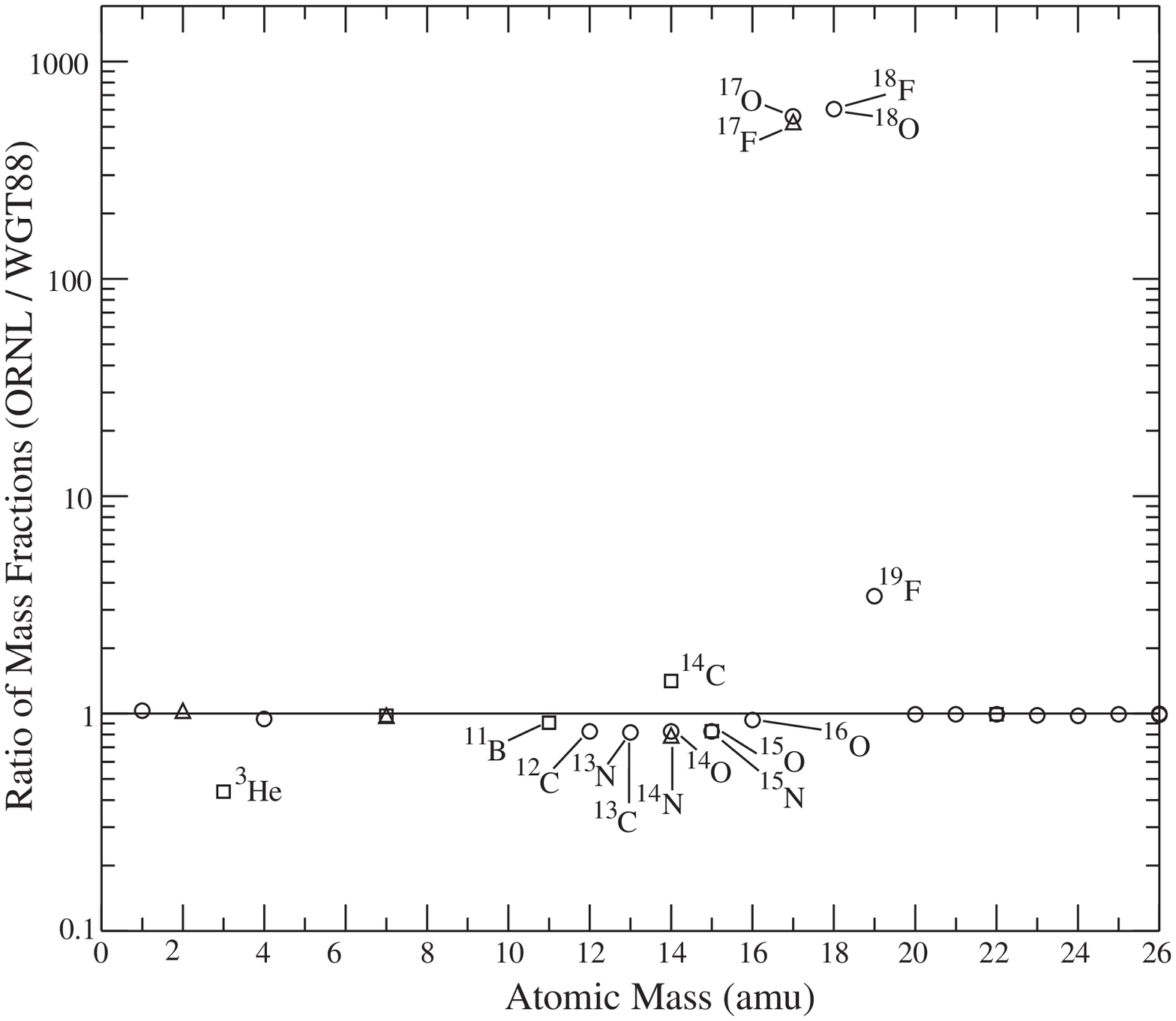} 
    \caption{The ratio of mass fractions (ORNL/WGT88) plotted as a function   
    of the nuclide mass for the hottest zone of a 1.25 $\msun$ WD nova. The 
    ORNL rate changes the mass fractions of some nuclei by up to a 
    factor of 600. The different symbols represent the same abundance 
    levels as in Figure~\ref{fig:ratio125}. 
    \label{fig:ratio125z1}} 
\end{figure}

Comparisons between the mass fractions calculated with the ORNL and WGT88
rates show the largest changes for the nuclides $^{18}$F (ORNL/WGT88 mass
fraction ratio = 2.08), $^{18}$O (2.00), $^{17}$O (1.92), $^{17}$F (1.77),
$^{14}$C (1.39), $^{14}$N (0.776), $^{14}$O (0.745), $^{13}$C (0.702),
$^{15}$O (0.629), $^{15}$N (0.542), and $^{19}$F (0.433), when all zones of
the nova were considered (Figure~\ref{fig:ratio125}). A comparison of the
mass fractions from the hottest zone (Figure~\ref{fig:ratio125z1}), which
contains $\sim 16\%$ of the total mass of the envelope and where the
largest number of the nuclear transmutations occur, shows the largest
changes for $^{18}$F (ORNL/WGT88 mass fraction ratio = 604), $^{18}$O
(604), $^{17}$O (559), $^{17}$F (527), $^{19}$F (3.5), and $^{3}$He (0.43).

The simple expectation is that the network with the faster
$^{17}$F(p,$\gamma$)$^{18}$Ne rate would produce more $^{18}$F because this
nuclide is the direct decay product of $^{18}$Ne. This is the case for the
cooler outer regions of these nova, where the lower temperatures do not
allow a high probability for $^{18}$F destruction via 
$^{18}$F(p,$\alpha$)$^{15}$O.  Thus the early surplus of $^{18}$F
production caused by the faster $^{17}$F(p,$\gamma$)$^{18}$Ne rate is
maintained as the envelope cools. This is not, however, the case for the
hotter inner zones where most of the nucleosynthesis occurs. Calculations
with the faster $^{17}$F(p,$\gamma$)$^{18}$Ne rate do produce an early
surplus of $^{18}$F in these zones, but at times when the temperatures are
high. This allows the $^{18}$F(p,$\alpha$)$^{15}$O reaction to destroy this
surplus, though convective mixing to cooler layers may preserve some of
this surplus.

\begin{figure}
    \includegraphics[width=\linewidth] {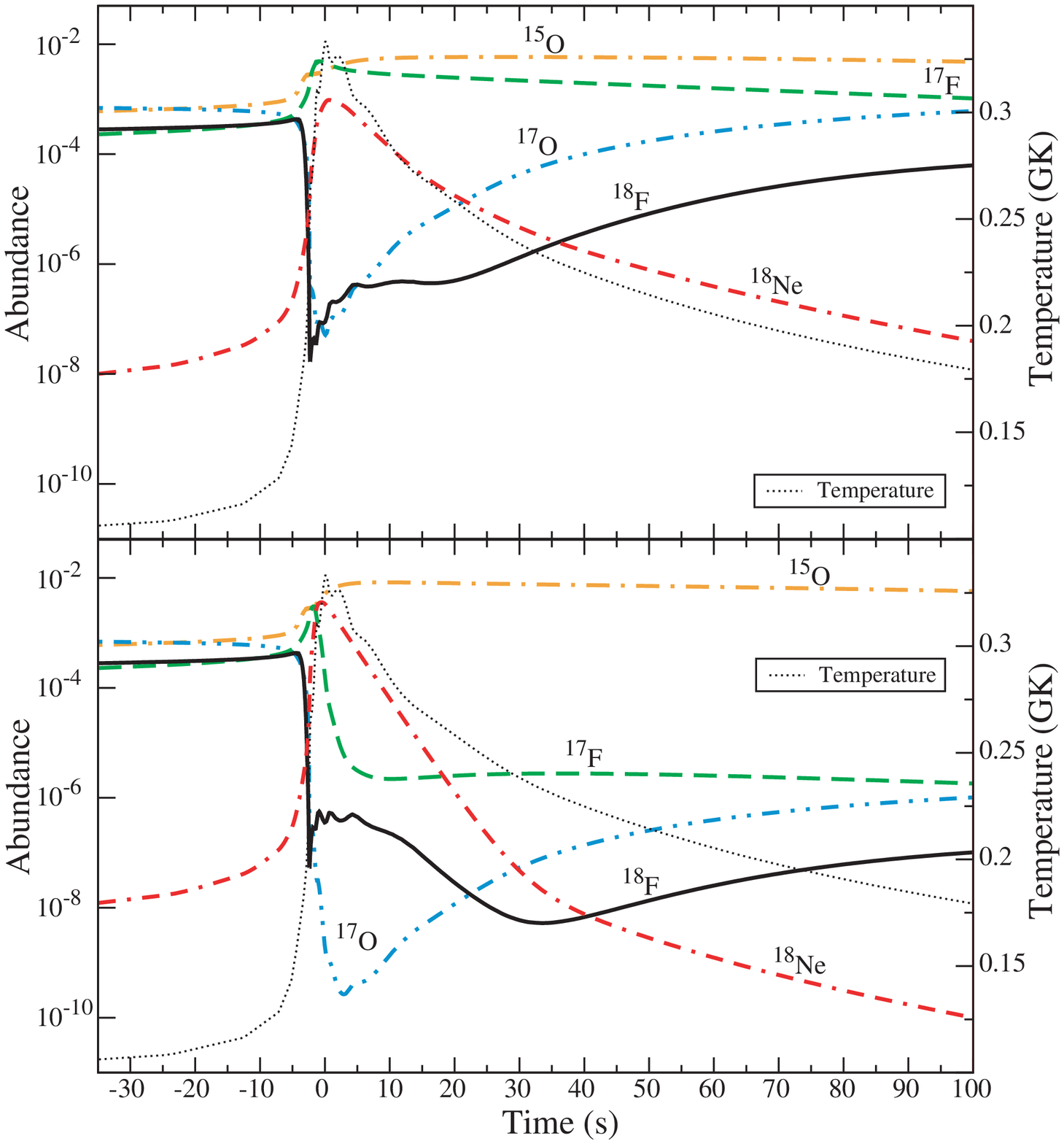} 
    \caption{ Abundances plotted as a function of time for the hottest 
    zone of the 1.25 $\msun$ WD nova. The abundances calculated by the 
    network with the new, slower ORNL $^{17}$F(p,$\gamma$)$^{18}$Ne rate 
    are shown in (a) and the abundances calculated with the older, 
    faster $^{17}$F(p,$\gamma$)$^{18}$Ne WGT88 rate are shown in (b). 
    The temperature history for this zone is also plotted.   
    \label{fig:yvst}}
\end{figure}

Models with the slower $\mathrm{^{17}F(p,\gamma)^{18}Ne}$ rates produce
more $^{18}$F in the hotter, inner zones (via $\mathrm{^{17}F (\beta^+\nu)
^{17}O(p,\gamma)^{18}F}$, leading to a larger amount of $^{18}$F when
averaged over the entire envelope.  This can be seen in
Figure~\ref{fig:yvst} which shows the time evolution of several abundances
in the hotter, innermost ejecta zone of the 1.25 $\msun$ WD nova. There are
two reaction paths that lead from $^{17}$F to $^{18}$F: $\mathrm{^{17}F
(p,\gamma) ^{18}Ne (\beta^+\nu) ^{18}F}$ and $\mathrm{^{17}F (\beta^+\nu)
^{17}O(p,\gamma)^{18}F}$. Figure~\ref{fig:yvst} shows that for 5 seconds
after the time of peak temperature, the abundance of $^{18}$F is larger for
the WGT88 case which has the faster $^{17}$F(p,$\gamma$)$^{18}$Ne rate.
This case also produces more $^{18}$Ne, demonstrating that the production
of $^{18}$F is dominated by $\mathrm{^{17}F (p,\gamma) ^{18}Ne (\beta^+\nu)
^{18}F}$ during this period. However, the highest temperatures of the nova
occur during this period, leading to further processing of the
freshly-synthesized $^{18}$F.  The faster $^{17}$F(p,$\gamma$)$^{18}$Ne
rate does allow more transmutations from $^{17}$F to $^{18}$F, but the
$^{18}$F is soon destroyed, primarily via the
$^{18}$F(p,$\alpha$)$^{15}$O reaction. The network with the faster WGT88
$\mathrm{^{17}F(p,\gamma)^{18}Ne}$ rate shows 62\% more $^{15}$O production
compared to  the network with the slower rate (ORNL case) 5 seconds after
peak.

The final abundance of $^{17}$F and $^{17}$O produced in the ORNL case is
greater than the abundance of $^{17}$F and $^{17}$O produced in the WGT88
case because the slower $^{17}$F(p,$\gamma$)$^{18}$Ne ORNL rate allows more
$^{17}$F to survive and decay to $^{17}$O. Beginning $\sim 20 - 30$ seconds
after the time of peak temperature, the abundance of $^{17}$O becomes
greater than the abundance of $^{18}$Ne, allowing production of $^{18}$F to
be dominated by the reaction sequence $\mathrm{^{17}F (\beta^+\nu) ^{17}O
(p,\gamma)^{18}F}$. Graphically this can be seen in Figure~\ref{fig:yvst} 
by noting that the time evolution of the $^{18}$F and $^{17}$O abundances
are parallel after the $^{18}$Ne abundance drops below the $^{17}$O
abundance. This figure also shows that the ORNL case starts this period
with 300 times more $^{17}$F and $^{17}$O than the WGT88 case. In this
period, there is little change in the abundance of $^{15}$O because the
temperature has dropped, inhibiting the destruction of  $^{18}$F via the
$^{18}$F(p,$\alpha$)$^{15}$O reaction. Thus the slower
$^{17}$F(p,$\gamma$)$^{18}$Ne rate produces a larger final abundance of
$^{18}$F than the faster rate because the delayed production of $^{18}$F to
the cooler post-peak temperatures allows $^{18}$F to survive as $^{18}$F
and its decay product $^{18}$O.

The SF98 and GAM91 $^{17}$F(p,$\gamma$)$^{18}$Ne reaction rates are,
respectively, only 12\% and 8\% smaller than the ORNL rate at the peak
temperatures of the 1.25 $\msun$ WD nova. Our calculations show that there
are only small differences in the mass fractions calculated with networks
containing these three rates. The largest changes, when all zone are
considered, are between the SF98 and ORNL cases: -1.5\% $^{14}$O, -1.3\%
$^{18}$F, -1.3\% $^{18}$O, -1.0\% $^{17}$F, and +1.1\% $^{19}$F when the
ORNL rate is used. For the hottest zone, calculations with the ORNL rate
result in 6.5\% less $^{17}$F, 3.5\% less $^{18}$F and $^{18}$O, 3.4\% less
$^{17}$O, 2.4\% less $^{17}$F, and 2.0\% more $^{14}$O compared to
calculation using the SF98 rate.  Comparison of calculation using the GAM91
and ORNL rates show changes of -0.7 \% $^{18}$F and $^{18}$O and +0.6\%
$^{19}$F in the ORNL case for the weighted sum of all zones in the 1.25
$\msun$ WD nova model.  For the hottest zone, calculations using the ORNL
rate result in 2.0\% less $^{17,18}$F, and $^{17,18}$O, and 1.4\% less
$^{19}$F than those using the GAM91 rate.

The zone by zone analysis reveals that the SF98 case produces more $^{18}$F
than the ORNL case for the first three zones, and the GAM91 case produces
more $^{18}$F than ORNL for only the first two zones. The ORNL
$^{17}$F(p,$\gamma$)$^{18}$Ne rate is slightly faster than the SF98 or
GAM91 rates and there is only a small difference in the depletion of
$^{17}$F between the three cases. The slower SF98 and GAM91 $\mathrm{^{17}F
(p,\gamma)^{18}Ne}$ rates allow even more $^{18}$F production to occur
later, and therefore, at lower temperatures than $^{18}$F production in the
ORNL case.

In all three cases, the network with the faster $\mathrm{^{17}F (p,\gamma)
^{18}Ne}$ rate produces more $^{18}$F in the outer zones because the lower
temperatures in the outer zones do not allow for significant destruction of
$^{18}$F via $\mathrm{^{18}F(p,\alpha)^{15}O}$. Thus the early surplus of
$^{18}$F in these zones caused by the faster $\mathrm{^{17}F (p,\gamma)
^{18}Ne}$ rate is maintained as the envelope cools. This difference in
behaviour between the inner and outer zones of an individual model shows
the importance of considering nucleosynthesis (and mixing) throughout the
nova envelope. If the hottest zone alone was considered, an incorrect
conclusion would have been drawn regarding the changes in the abundance of
$^{18}$F.
    
\section{Results for a 1.35 $\msun$ ONeMg WD Nova}

\begin{figure}
    \includegraphics[width=\linewidth] {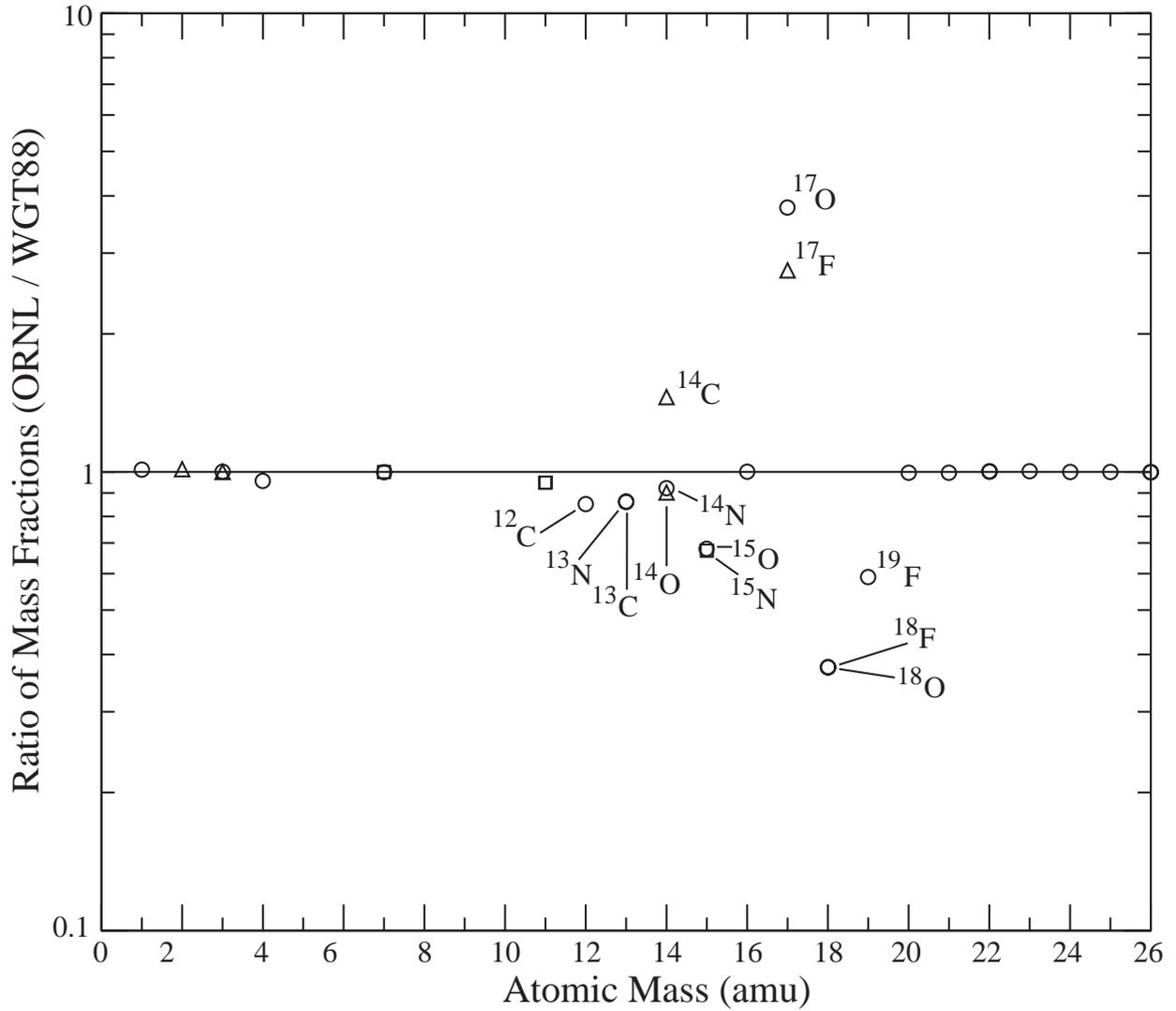} 
    \caption{ The ratio of mass fractions (ORNL/WGT88) plotted against 
    nuclide mass for the entire envelope of a 1.35 $\msun$ WD nova. The 
    ORNL rate changes the mass fractions of some nuclei by up to a 
    factor of 3.8. The different symbols represent the same abundance 
    levels as in Figure 2. 
    \label{fig:ratio135}}
\end{figure} 

\begin{figure}
    \includegraphics[width=\linewidth] {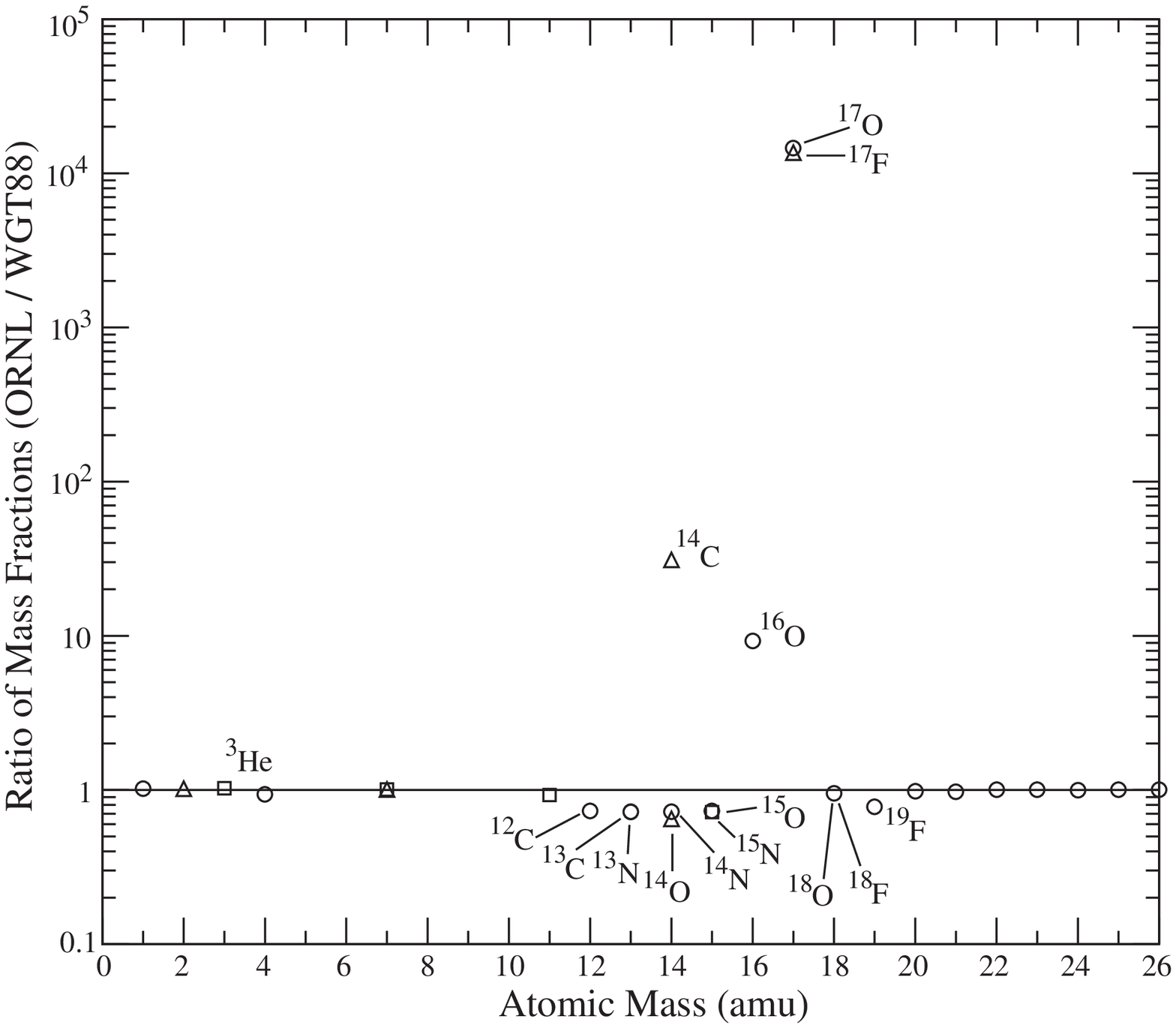} 
    \caption{ The ratio of mass fractions (ORNL/WGT88) plotted against 
    nuclide mass for the third hottest zone of the 1.35 $\msun$ WD nova. The 
    ORNL rate changes the mass fractions of some nuclei by up to a 
    factor of 15000. The different symbols represent the same abundance 
    levels as in Figure~\ref{fig:ratio125}. 
    \label{fig:ratio135z3} } 
\end{figure}

The hotter 1.35 $\msun$ WD nova simulation showed the greatest variation in
the mass fraction patterns produced by the four rates. Comparisons between
the mass fractions calculated with the ORNL and WGT88 rates show the
largest changes for $^{17}$O (ORNL/WGT88 mass fraction = 3.8), $^{17}$F
(2.8), $^{18}$F (0.36), and $^{18}$O (0.38) when the entire exploding
envelope was considered.  A more complete sample of abundance ratios are
shown in Figures~\ref{fig:ratio135}~\&~\ref{fig:ratio135z3}.

In the 1.35 $\msun$ model, the largest differences in nuclear production
between the ORNL and WGT88 cases were exhibited in the third hottest zone 
because the WGT88 rate differs more from the ORNL rate at 0.437 GK, the
peak temperature in this zone, than it does at 0.457 GK, the peak
temperature in the hottest zone.  Comparison of mass fractions synthesized
in the third hottest zone shows that using the WGT88 and ORNL
$\mathrm{^{17}F (p,\gamma) ^{18}Ne}$ rates results in very large
differences in the mass fractions of $^{17}$F (ORNL/WGT88 mass fraction
ratios = 15,000) and $^{17}$O (14,000). As Figure~\ref{fig:ratio135z3}
illustrates, the ORNL rate also produced 31 times more $^{14}$C and 9 times
more $^{16}$O than the WGT88 rate in this zone.  However, the production of
$^{18}$F and $^{18}$O was slightly reduced.

In the case of $^{18}$F, a zone by zone analysis of the 1.35 $\msun$ WD
nova shows that only the innermost zone produces more $^{18}$F when the
ORNL rate is used in place of the WGT88 rate. The 1.35 $\msun$ WD is the
most violent of the three novae considered here, reaching the highest peak
temperature, but also expanding, and therefore cooling, the fastest of the
three novae. The temperature in the hottest zone drops from .457 GK at peak
to 0.01 GK in just 60 seconds compared to the 307 seconds that the hottest
zone in the 1.25 $\msun$ WD model takes to drop from a peak temperature of
0.333 GK to 0.01 GK. As was the case for the innermost zone of the 1.25
$\msun$ WD nova, the ORNL case produced more $^{18}$F than the WGT88 case
because the WGT88 case produced its $^{18}$F while temperatures remained
high enough for the $^{18}$F to be destroyed via
$^{18}$F(p,$\alpha$)$^{15}$O. For the rest of the ejecta, the early excess
in $^{18}$Ne and $^{18}$F production using the WGT88 rate is maintained
because of the very rapid drop in temperature, limiting the later $^{18}$F
destruction via the (p,$\alpha$) reaction. The final result is a smaller
abundance of $^{18}$F from the ORNL rate than from the WGT88 rate
(Figure~\ref{fig:ratio135}).

As with the 1.25 $\msun$ WD nova, there were much smaller differences in
mass fractions produced using the ORNL rate and the SF98 and GAM91 rates. 
In the SF98 and GAM91 cases the hottest zone showed the largest differences
from the ORNL case because the SF98 and GAM91 rates differ more from the
ORNL rate at temperatures above 0.47 GK rather than below it (see
Figure~\ref{fig:rateratio}). In the comparison with the SF98 case, when the
entire exploding envelope was considered, the ORNL case resulted in 16\%
less $^{14}$C, 5\% more $^{18}$O and $^{18}$F, 5\% less $^{17}$O, and 3\%
less $^{17}$F; all other changes were smaller. For the hottest zone, the
largest change was a 66\% decrease in $^{17}$F and $^{17}$O using the ORNL
rate rather than the SF98 rate. In comparison with the GAM91 case, when the
entire exploding envelope was considered, the ORNL case produced 3\% more
$^{18}$F and $^{18}$O, 2\% more $^{19}$F, 9\% less $^{14}$O, 3\% less
$^{17}$O, and 2\% less $^{17}$F. The ORNL case produced about 48\% less
$^{17}$F, and $^{17}$O, and 12\% less $^{16}$O than the GAM91 case in the
hottest zone; all other changes were smaller.

\section {Results for a 1.00 $\msun$ CO WD Nova}

Because of the lower peak temperatures, the 1.00 $\msun$ WD nova
simulations showed the least variation in the abundance patterns produced
by the four rates. The largest variation was in the mass fraction of
$^{18}$Ne: 21\% more $^{18}$Ne was produced using the WGT88 rate when
compared to the ORNL case, both for the hottest zone and for the weighted
sum of all zones. There was less than 0.3\% change in the mass fractions of
all other isotopes when comparing calculations using the ORNL and WGT88
rates. The ORNL rate resulted in 1.8\% more $^{18}$Ne than the SF98 rate
and 1.0\% more $^{18}$Ne than the GAM91 rate in both the hottest zone and
weighted average of all zones. There was less than 0.4\% variation in the
mass fractions of all other isotopes between the different rate cases.

\section{Comparison with Observations and Other Studies}

Novae introduce large quantities of gas and dust into the interstellar
medium, and this material can be preserved in meteorites that fall to
Earth. Five meteoritic dust grain inclusions have recently been shown to
have  isotopic signatures matching those predicted for ONeMg nova ejecta
\citep{AGNZ01,AZJH01}. Compared to solar abundances, these grains are
characterized by low $^{12}$C /$^{13}$C and $^{14}$N /$^{15}$N abundance
ratios, a high $^{26}$Al/$^{27}$Al abundance ratio, and large excesses of
$^{30}$Si. Theoretical nucleosynthesis calculations show that novae ejecta
are the only stellar sources that match this isotopic composition
\citep{AGNZ01,AZJH01}.

Our calculated abundance ratios compare relatively well with those measured
in the grains, as well as those in other nucleosynthesis calculations.  The
isotopic ratio of $^{12}$C/$^{13}$C for solar composition is 89.9
\citep{AnGr89}, while for the five meteoritic dust grains this ratio ranges
from 4 to 9.4 \citep{AGNZ01,AZJH01}. Our 1.25 $\msun$ and 1.35 $\msun$
ONeMg WD nova models predict $^{12}$C/$^{13}$C ratios of 3.2 and 5.6
respectively, while nucleosynthesis calculations from other studies predict
ratios between 0.3 and 3 for ONeMg nova ejecta
\citep{AGNZ01,AZJH01,JoCH99}. The lower limit of the $^{26}$Al/$^{27}$Al
ratio was determined to be 0.8 for one of the meteoric dust grains and the
ratio for another was 0.0114. Our study predicted that the ratio would be
0.15 and 0.03 for the 1.25 $\msun$ and 1.35 $\msun$ ONeMg WD nova models,
respectively. Other nucleosynthesis studies predicted a ratio between 0.07
and 0.7 \citep{AGNZ01,AZJH01,JoCH99}. The agreement is not universally
good, however. The $^{14}$N/$^{15}$N  ratios are between 5.22 and 19.7 for
4 of the 5 meteoritic grains, while our study predicts ratios of 0.06 and
0.51 for the 1.25 and 1.35 $\msun$ ONeMg WD nova models, respectively. By
comparison, the solar composition ratio for $^{14}$N/$^{15}$N is 270
\citep{AnGr89}, so both the meteorite value and our prediction show a
significant depletion in $^{14}$N compared to solar. Other nucleosynthesis
studies predict a nitrogen isotopic ratio overlapping with ours -- between
0.1 and 10 \citep{AGNZ01,AZJH01}. These comparisons show that the isotopic
ratios determined from our 1.25 and 1.35 $\msun$ WD novae nucleosynthesis
calculations qualitatively match those measured in grains thought to
originate in novae, but further study of the impact of nuclear
uncertainties on nova nucleosynthesis is certainly warranted.

Novae are thought to overproduce $^{17}$O and $^{18}$O abundances relative
to solar, and the new $^{17}$F(p,$\gamma$)$^{18}$Ne reaction rate has a
strong influence on the predicted synthesis of these isotopes. The new ORNL
$^{17}$F(p,$\gamma$)$^{18}$Ne rate significantly changes the predictions of
these isotopic abundances compared to the rate based on the  
resonance parameters in \citep{WiGT88}.  Unfortunately, no oxide grains
of putative novae origin have yet been identified. The discovery and
analysis of oxide grains from novae would provide an important constaint on
models of nova nucleosynthesis.

\section{Summary}

The abundances of certain nuclides synthesized in nova explosions on 1.25
and  1.35 $\msun$ ONeMg white dwarfs have been shown to depend strongly on
the rate of the $^{17}$F(p,$\gamma$)$^{18}$Ne reaction.  A new, slower rate
for this reaction, based on a measurement with a $^{17}$F radioactive beam
at ORNL, significantly changes the abundance predictions for $^{17}$O and
$^{17}$F synthesized in the hottest zones of the explosion on a 1.35
$\msun$ WD by up to factor of 14,000 compared to predictions using a
$^{17}$F(p,$\gamma$)$^{18}$Ne reaction rate based on resonance parameters 
employed in the most widely-used rate library. The new rate also changes the production of several isotopes ($^{12,13,14}$C, $^{13,14,15}$N, $^{15,17,18}$O, and $^{17,18,19}$F) by as much as a factor of 3.7 when entire ejected envelope is considered. The calculations for a 1.25 $\msun$ WD nova show that this new rate changes the abundances of $^{17,18}$O and $^{17,18}$F synthesized in the hottest zones up to a factor of 600 compared to some previous estimates, and changes the abundances of $^{12,13,14}$C, $^{13,14,15}$N, $^{15,17,18}$O, and $^{17,18,19}$F by up to a factor of 2.1 when averaged over the entire exploding envelope.  For 1.00 $\msun$ CO WD nova nucleosynthesis calculations, the peak temperatures are
low enough that the values of the $^{17}$F(p,$\gamma$)$^{18}$Ne reaction
rate differ only slightly, causing only small differences in the
nucleosynthesis.

Regarding the production of the important, long-lived radionuclide
$^{18}$F, we find that the production is increased in the hottest regions
of the nova by the slower ORNL $^{17}$F(p,$\gamma$)$^{18}$Ne rate. A faster
$^{17}$F(p,$\gamma$)$^{18}$Ne rate creates $^{18}$F (from decay of
$^{18}$Ne) sooner after the peak of the outburst and therefore at higher
temperatures -- where it is more likely to be destroyed by
$^{18}$F(p,$\alpha$)$^{15}$O. The slower $^{17}$F(p,$\gamma$)$^{18}$Ne rate
slows the production of  $^{18}$F, creating more of it at a lower
temperature, where it is more likely to survive as a mass 18 isotope.  This
effect does not, however, carry over to the outer zones of the explosion,
because the overall lower temperatures of these zones limits the post-peak
destruction of freshly-synthesized $^{18}$F. If only the innermost zones
were considered, an incorrect conclusion would be drawn regarding the
change in the synthesis of $^{18}$F -- showing the importance of
considering the nucleosynthesis throughout the entire nova envelope.  Our
predicted isotopic ratios qualitatively agree with those measured in grains
from nova ejecta, as well as with other nova nucleosynthesis calculations.

Our study shows the importance of using the best nuclear physics input for
calculations of the nucleosynthesis occurring in nova explosions. Even
though the reaction network involves hundreds of highly-coupled reactions,
an improvement to the estimate of an individual reaction rate can  make a
significant changes in element synthesis predictions, some of which may be
directly observable through $\gamma$-ray astronomy or the study of presolar
grains. Our results also show the strong sensitivity that nucleosynthesis 
calculations have to hydrodynamic profiles. With improved nuclear physics input, nucleosynthesis calculations can begin to differentiate between competing hydrodynamic simulations of nova outbursts. 

\acknowledgements The authors acknowledge helpful comments on the
manuscript from R. Kozub and the anonymous referee. Oak Ridge National
Laboratory is managed by UT-Battelle, LLC, for the U.S. Department of
Energy under contract DE-AC05-00OR22725. This work has been partly
supported by NASA under contract NAG5-8405, by the National Science
Foundation under contract AST-9877130, and by funds from the Joint
Institute for Heavy Ion Research. SS acknowledges partial support from NSF
and NASA grants to Arizona State Univ.

%\bibliographystyle{apj} 
%\bibliography{apjmnemonic,nova,hix}

\end{document}